\documentclass[final,english]{article}
\usepackage{ifdraft}
\ifdraft{
   \usepackage{showidx,showlabels,showkeys,drftcite}
}{}
\usepackage[utf8]{inputenc}
\usepackage[english]{babel}
\usepackage{amsmath}
\usepackage{SIunits}
\usepackage{enumerate}
\usepackage{tabularx}
\usepackage{dcolumn}
\usepackage[final]{listings}
\usepackage[babel]{csquotes}

\usepackage{hyperref}

\newcommand\enbrace[1]{\ensuremath{\left(#1\right)}}

\newcommand\comment[1]{\relax}
\makeatletter
\newcommand{\set}[2][\relax]{%
   \ifx#1\relax \ensuremath{\left\lbrace#2\right\rbrace}
   \else \ensuremath{%
     \setbox0=\hbox{\ensuremath{#2}}%
     \dimen@\ht0%
     \advance\dimen@ by \dp0%
     \left\lbrace\left.#1\rule[-\dp0]{0pt}{\dimen@}\right|#2\right\rbrace}%
   \fi%
}
\makeatother

\begin{document}
\title{Analysis of tubes filled with charged electron gas}
\author{Dr. Stefan Karrmann\\[1ex]
   \small Stuttgart\\
   \small Germany}
\date{2011-02-23}
\maketitle

\begin{abstract} \noindent
We show that tubes filled with electron gas, as presented by
A.~Bolonkin, are not possible with current materials.
First, the pressure of the charges on the outer surface cancel
almost all of the electrostatic pressure of the inner electrons.
Second, due to the mutually repulsion most of the electrons are in
the outmost shell of the tube and not individually free.
\end{abstract}

\tableofcontents

\section{Introduction}
Tubes filled with electron gas, as described in \cite{AB-Tube}, are
very versatile devices.
First, they are very light -- almost as light as vacuum, i.e. much lighter
than hydrogen filled tubes. This predestined them for e.g. ballons
and high altitude platforms.
Second, they are almost perfect conductors at huge range of temperatures, in
particular at room temperature (circa \unit{300}\kelvin).
Thus, a detailed analysis of its properties is necessary.
The sketch of the construction is given in \cite{AB-Tube}.
Here we present additional arguments which show that we cannot
construct such tubes at a macroscopic scale with known materials.

We provide mainly two arguments to the analysis of the tubes. 
First, we pay attention to the forces which act on the positivly
charged outer surface.
Second, we study the behaviour of electrons in the interior of the
tube and compare this to the described behaviour of the electon gas
in \cite{AB-Tube}:
\blockquote[{\cite[page 2]{AB-Tube}}]{%
   The applications of electron gas are based on one little-known fact --
   the electrons located within a cylindrical tube having a positively
   charged cover (envelope) are in neutral-charge conditions -- the total
   attractive force of the positive envelope plus negative contents equals
   zero. That means the electrons do not adhere to positive charged tube
   cover. They will freely fly into an AB-Tube.
   }

In section~\ref{sec:design} we describe the construction of the tube
as in \cite{AB-Tube} and name its parameters.
In section~\ref{sec:gauss} we use Gauss's law to analyse the
electrical force and particle acceleration of the electrons inside the tube.
In section~\ref{sec:chargeTube} we describe what happens if we
charge the tube interior.
Then we are ready to calculate all actual forces of the tube in
section~\ref{sec:capacitor}.
Finally, we give additional arguments to support the behaviour of
the electron gas in section~\ref{sec:arguments}.

\section{Design of the tube}
\label{sec:design}
To describe the tube we use cylinder coordinates. The tubes
centerline rests on the z-axis of the coordinate system.
The interior of the tube is empty and is surrounded by a
non-conducting envelope. This envelope has a conducting outer
surface. The parameters and their units of this tube are:\\
\begin{tabular}[t]{lp{10cm}}
\(d\) & the envelope thickness [\(\meter\)]\\
\(l\) & the length of the tube [\(\meter\)]\\
\(S_1\) & the inner surface of the envelope\\
\(S_2\) & the outer surface of the envelope\\
\(r_1\) & the radius of the inner surface \(S_1\) [\(\meter\)]\\
\(r_2\) & the radius of the outer surface \(S_2\) [\(\meter\)]\\
\(A_i\) & the area of the surface \(S_i\) [\(\meter^2\)]\\
\(\tau_1\) & the one dimensional charge density of all charges on
   \(S_1\) or inside the tube
   [\(\coulomb\per\meter\)]\\
\(\tau_2\) & the one dimensional charge density of all charges on \(S_2\)
   [\(\coulomb\per\meter\)]\\
\(\epsilon\) & the permittivity \(\epsilon_r\epsilon_0\) [\(\farad\per\meter\)]\\
\(\epsilon_r\) & the relative permittivity of the envelope [1]\\
\(E_i\) & the electrical field generated by the charges
        \(\tau_i\) on \(S_i\) in the radial direction
        [\(\newton\per\coulomb\)]
\end{tabular}\\
We have the relations \(r_2 = r_1 + d\), \(A_i = 2\pi r_i l\) by
the geometry of the tube.
We assume \(\tau_2=-\tau_1>0\) which is equivalent to
electrically neutrality of the whole tube.

\section{Gauss's law}
\label{sec:gauss}
In this section we apply Gauss's law, also known as Gauss's flux
theorem, to the tube. To simplify the calculation, we assume:
\begin{enumerate}
\item\label{enum:equi} The system is in a dynamical equilibrium.
\item\label{enum:len} The length \(l\) of the tube is big enough such that we can
neglegt the bottom and the top of the tube.
\item\label{enum:sym} The axial symmetry of the tube carries over to the charge
distribution.
\end{enumerate}
Assumption~\ref{enum:equi} is justified as electrons are quite
mobile and have low inertia.
We can approximate assumption~\ref{enum:len} as good as necessary
by construction of the tube.
Finally, assumption~\ref{enum:sym} is common in physics and is
justified here because of the homogenity along the axis as long as
we are far away from the ends of the tube which is given by
assumption~\ref{enum:len}. We could drop it by introducing the
position on the axis as a new parameter, but it complicates the
calculation without leading to new insights.

As of the symmetry we have a one dimensional charge density
\(\tau(r)\leq0\) which depends only on the distance to the central
axis. It describes the charge per length in a sub-tube of radius
\(r\) whichs axis is identical with the axis of the tube.
As we neglect the end sides of the tube we have that the direction
of the electrical field \(\mathbf E\) is radial.
Now we can apply Gauss's law the the Volume \(V(r)\) of a small
finite  cylinder whichs axis is on the axis of the tube, radius
\(r\), and length \(L\).

\begin{align}
&&\oint_{\partial V(r)} \mathbf{E} \cdot \mathrm{d}\mathbf{A}
& = \frac{Q(V(r))}{\varepsilon_0}
\\
&\iff&
E(r) 2\pi r L
& = \frac{\tau(r)L}{\varepsilon_0}
\label{eq:gaussRadial}&
\\
&\iff&
E(r)
& = \frac{\tau(r)}{2\pi r\varepsilon_0}
\label{eq:gaussE}
\end{align}

This means that an electron at distance \(r\) from the axis reacts
to the field \(\mathbf E(r)\) such that it accelerates radially by
\begin{align}
\label{eq:electronAccel}
a(r) & = E(r) q_e/m_e\geq0 \\
\text{where}\quad q_e
&=\unit{-1.602 176 487(40)\cdot10^{-19}}\coulomb&&
\text{(charge
of an electon)} \\
m_e &=\unit{9.109 382 15(45)\cdot10^{-31}}\kilogram\footnotemark
&&\text{(mass of an resting election)}
\end{align}
\footnotetext{Figures are taken from \cite{PDG}.}

Obviously the electrons in the tube are not free as long as there
are electrons nearer to the axis.
They repel each other such that they accelerate
radially by \(a(r)\) away from the axis.
If the electrons are not
symmetrically distributed the electrical field would be more
complicated but not zero everywhere.

Thus, there is no such thing as an electron gas which consists of
individually free electrons.

\section{Charging of the tube}
\label{sec:chargeTube}
If we charge the tube interior, e.g. by a hot cathode or by field
electron emission, the electrons will repel each other following
equation~\eqref{eq:electronAccel}.
Sooner or later they hit the non conducting inner surface \(S_1\) of
the tube.
There, some may be elastically or inelastically reflected other will
adhere to the insulater as static electricity.
The reflected electrons are accelarated according to
equation~\eqref{eq:electronAccel} and will hit \(S_1\) again and
again until they adhere there as static electricity.

The charge distribution of the static electricity on \(S_1\) will be
uniform mainly because of two reasons.
First, the electrons have higer propability to hit regions with
(temporarily) lower charge, i.e. electron density.
Second, every real insulater has a finite electric resistance such
that any non-uniform charge distribution adjust over time to the
uniform distribution by small electric currents.

Finally, all electrons which have been injected into the tube ends
as static electricity of the insulater at the inner surface \(S_1\)
and that they are uniformly distributed.

Nota bene, an uniformly distributed charge on \(S_1\) does not
generate an electrical field inside the tube because of
Gauss's law and the symmetry resulting in equation~\pageref{eq:gaussRadial}.

\section{Capacitor model}
\label{sec:capacitor}
Given that the charges inside the tube adhere uniformly distributed
to the inner surface \(S_1\) we can calculate the forces which acts
on the tube.
Since the outer surface \(S_2\) is conducting and positivly charged
such that the total charge is \unit0\coulomb{} the tube is a
cylinder capacitor.

In \cite{AB-Tube} the analysis focuses on the force which acts at
the inner surface \(S_1\) and that the force is generated by the
electrical field as electrical pressure, c.f. \cite[equation (1) on
page 6]{AB-Tube}.

In the discussion in \cite{AB-Tube} all forces are ignored which act on the
positive charges at the outer surface \(S_2\) of the envelope. To understand the
situation and the forces at the envelope, we first analyse the effects of
the electrical field next to the electrically insulating envelope.

We use the result~\eqref{eq:gaussE} to calculate \(E_i\). Since we
are interested in the electrical field next to the surfaces \(S_i\)
but in the material of the envelope we use the limit to get the
result.
\begin{align}
E_1 &= \lim_{h\to0+} E(r_1+h)
     = \frac{\tau_1}{2\pi r_1\varepsilon}\\
E_2 &= \lim_{h\to0+} E(r_2-h)
     = \frac{\tau_1}{2\pi r_2\varepsilon}
\end{align}

The pressure of the electric field is given by 
\begin{align}
   p_i  & = \frac\epsilon2 E_i^2
\end{align}
The direction of the force is in each case into the envelope, i.e.
\(p_1\) is outward radial pressure and \(p_2\) is inward radial
pressure.
This is the case since if we would move the charges each in its
direction we would decrease the total energy of the system.
Equivalently, we would reduce the volume which contains the
electrical field without increasing it due to Gauss's law. The
energy of the electric field, whichs energy density is
\(2E^2/\epsilon\), in the volume difference would be set free.

Now, that we know the direction of the pressure we can calculate the
total force \(F_\text{tot}\) which acts on the envelope in the
radial direction by:
\begin{equation}
\begin{aligned}
   F_\text{tot}
   & = F_2 + F_1
   = p_2 A_2 + p_1 A_1
   = \pi \epsilon l \enbrace{r_1 E_1^2 - r_2 E_2^2} \\
   & = \frac{l \tau_2^2}{4\pi\epsilon} \enbrace{\frac1{r_1} - \frac1{r_2}}
   = \frac{l \tau_2^2}{4\pi\epsilon} \frac{r_2 - r_1}{r_2 r_1}\\
   & = \frac{l \tau_2^2}{4\pi\epsilon} \frac d{r_1^2+r_1 d}\\
\end{aligned}
\end{equation}
We see, that for infinite thin envelopes, i.e. \(d\to0\), the total
force is zero \(F_\text{tot} = 0\).
For \(d\ll r_1\) we can approximate
\begin{equation}
   F_\text{tot} \approx \frac{l \tau_2^2}{4\pi\epsilon} \frac
   d{r_1^2}.
\end{equation}
To overcome the atmospheric pressure \(p_\text{air}\) we need a total force
\(F_\text{tot}\) such that:
\newcommand{\absVal}[1]{\left|#1\right|}
\begin{equation}
   \label{geq:air}
   F_\text{tot} \geq p_\text{air} A_2
   = 2 \pi r_2 l p_\text{air}
\end{equation}
On the other hand we must limit the electrical field \(E_i\) by the
dielectric strength \(E_\text{max}\) of real materials, c.f. appendix ~\ref{ssec:material}.
As the field \(\mathbf E\) is stronger at the inner surface \(S_1\) we must obey:
\begin{equation}
   \label{leq:E-max}
   E_1 \leq E_\text{max}
\end{equation}
We simplify the two conditions~\eqref{geq:air}, \eqref{leq:E-max}. The
result turns out to be:
\begin{align}
   \comment{
   F_\text{tot} & = \frac{l \tau_2^2}{4\pi\epsilon} \frac d{r_1^2+r_1 d}\\
   F_\text{tot} & \geq 2 \pi r_2 l p_\text{air} \\
   \frac{l \tau_2^2}{4\pi\epsilon} \frac d{r_1^2+r_1 d} & \geq 2 \pi r_2 l p_\text{air} \\
   \frac{\tau_2^2}{8\pi^2\epsilon} \frac d{r_2(r_1^2+r_1 d)} & \geq p_\text{air} \\
   \frac{\tau_2^2}{8\pi^2\epsilon} \frac d{p_\text{air}} & \geq r_2^2 r_1 \\
   \tau_2 & \geq \sqrt{\frac{8\pi^2\epsilon p_\text{air} r_2^2 r_1}d} \\
   \tau_2 & \geq 2\pi r_2 \sqrt{\frac{2\epsilon p_\text{air} r_1}d} \\
   E_\text{max} & \geq \frac{\tau_2}{2\pi\epsilon r_1} \\
   E_\text{max} & \geq \frac{\tau_2}{2\pi\epsilon r_1}
   \geq \frac{2\pi r_2}{2\pi\epsilon r_1} \sqrt{\frac{2\epsilon p_\text{air} r_1}d} \\
   E_\text{max} & 
      \geq \frac{r_2}{\epsilon r_1} \sqrt{\frac{2\epsilon p_\text{air} r_1}d} \\
   E_\text{max} & 
      \geq r_2 \sqrt{\frac{2 p_\text{air}}{\epsilon r_1 d}} \\
   E_\text{max} & 
      \geq \sqrt{\frac{2 p_\text{air}}\epsilon}
           \frac{r_2}{\sqrt{r_1 d}} \\
   E_\text{max} \sqrt{\frac{\epsilon d}{2 p_\text{air}}}
      & \geq \frac{r_2}{\sqrt{r_1}} \\
   E_\text{max}^2 \frac{\epsilon d}{2 p_\text{air}}
      & \geq \frac{r_2^2}{r_1} \\
   E_\text{max}^2 \frac{\epsilon d}{2 p_\text{air}}
      & \geq \frac{(r2+d)^2}{r_1} \\
   }
   \frac{\epsilon d}{2 p_\text{air}} E_\text{max}^2
      & \geq \frac{r_1^2+2r_1d+d^2}{r_1}
      \operatorname*{\rightarrow}_{d\to0}r_1
\end{align}
For \(E_\text{max}=\unit{670}{\mega\volt\per\meter}\), \(\epsilon_r=2.3\)
(for Polyethylen) and \(d=\unit{1}{\micro\meter}\)
we obtain \(r_1 \approx \unit{45.11}{\micro\meter}\).
\comment{
2.3*8.854187817*10^-12*10^-6/(2*101325)*(670*10^6)^2
reorderd for bc -ql
2.3*8.854187817*10^-12*(670*10^6)^2*10^-6/(2*101325)
result of bc -ql
.00004511069968582778
}
The exact solution turns out to be
\begin{align}
   \comment{
   r_1 & = d \sqrt{\frac{(2-y)^2}{4} - 1} - d\frac{2-y}2\\
   y & = \frac{\epsilon E_\text{max}^2}{2 p_\text{air}}\\
   r_1 & = d \sqrt{\frac{(2-\frac{\epsilon E_\text{max}^2}{2 p_\text{air}})^2}{4} - 1} - d\frac{2-\frac{\epsilon E_\text{max}^2}{2 p_\text{air}}}2 \\
   r_1 & = d \sqrt{\frac{(2-\frac{\epsilon E_\text{max}^2}{2 p_\text{air}})^2}{4} - 1}
      - d\frac{4 p_\text{air}-\epsilon E_\text{max}^2}{4 p_\text{air}}\\
   r_1 & = d \sqrt{\frac{(4p_\text{air}-\epsilon E_\text{max}^2)^2}{(4p_\text{air})^2} - 1}
      - d\frac{4 p_\text{air}-\epsilon E_\text{max}^2}{4 p_\text{air}} \\
   r_1 & = \frac{d}{4 p_\text{air}}
         \enbrace{\sqrt{(4p_\text{air}-\epsilon E_\text{max}^2)^2- (4p_\text{air})^2}
            - (4 p_\text{air}-\epsilon E_\text{max}^2)} \\
   r_1 & = \frac{d}{4 p_\text{air}}
         \enbrace{\sqrt{\epsilon^2 E_\text{max}^4-8p_\text{air}\epsilon E_\text{max}^2}
            - (4 p_\text{air}-\epsilon E_\text{max}^2)} \\
   r_1 & = \frac{d}{4 p_\text{air}}
         \enbrace{\sqrt{\epsilon^2 E_\text{max}^4-8p_\text{air}\epsilon E_\text{max}^2}
            + (\epsilon E_\text{max}^2-4 p_\text{air})} \\
   }
   \label{eq:r2_exact}
   r_1 & = \frac{d}{4 p_\text{air}}
         \enbrace{\epsilon E_\text{max}^2-4 p_\text{air}
                 + \sqrt{\epsilon^2 E_\text{max}^4
                         - 8p_\text{air}\epsilon E_\text{max}^2} }
\end{align}
\comment{
epsilon_r = 2.3
epsilon_0 = 8.854187817*10^-12
epsilon = epsilon_r * epsilon_0
d = 10^-6
p_air = 101325
e_max = 670*10^6
"d/(4*p_air) * (epsilon * e_max^2 - 4*p_air + sqrt(epsilon^2*e_max^4-8*p_air*epsilon*e_max^2))"
d/(4*p_air) * (epsilon * e_max^2 - 4*p_air + sqrt(epsilon*e_max^2)*sqrt(epsilon*e_max^2-8*p_air))
.00004308749095311312

sqrt(1/.00004308749095311312)
152.34366429313152174735

sqrt(1/.00004308749095311312)*670*10^6
102070255076.39811957072450000000
}
which results in \(r_1 = \unit{43.08749}{\micro\meter}\). If we
increase the thickness of the envelope \(d\) the resulting radius \(r_1\)
increases exactly linearily.

If the pressure of the electric field must stabilize a macroscopic tube
with radius of \unit1{\meter} against standard atmospheric pressure
\(p_\text{air}\), we need new materials which have a dielectric
strength of about \unit{102}{\giga\volt\per\meter}, i.e. they must be more
than 150 times as strong as current existing materials, c.f.
\cite{WIKI-DE}.
Therefore, it is today impossible to create macroscopic tubes with thin
envelopes which are stabilized by electrostatic pressure.

Moreover, for an given envelope material the relative buoyancy is
proportional the the relative density if we neglect the end caps of
the tube. This implies that we do not obtain a ballon if we increase
the radius or length of the tube.
We give the proof in the remainder of this section.

Let \(m_\text{air}\) the mass of the displaced air [\(\kilogram\)],
and \(m_\text{env}\) the mass of the envelope of the tube [\(\kilogram\)].
Using \eqref{eq:r2_exact} we derive
\begin{align}
m_\text{air} & = \pi r_2^2 * l * \rho_\text{air}
               = \pi (r_1+d)^2 * l * \rho_\text{air}\\
m_\text{env} & = \pi (r_2^2-r_1^2) * l * \rho_\text{env}
               = \pi (2r_1d+d^2) * l * \rho_\text{env}\\
\frac{m_\text{env}}{m_\text{air}}
& = \frac{2r_1d+d^2}{r_1^2+2r_1d+d^2}
    \frac{\rho_\text{env}}{\rho_\text{air}}
  = \frac{2x+1}{x^2+2x+1}
    \frac{\rho_\text{env}}{\rho_\text{air}}\\
\text{where}\quad x & = \frac1{4 p_\text{air}}
         \enbrace{\epsilon E_\text{max}^2-4 p_\text{air}
                 + \sqrt{\epsilon^2 E_\text{max}^4
                         - 8p_\text{air}\epsilon E_\text{max}^2} }
\quad\text{by \eqref{eq:r2_exact}}\notag
\end{align}
The tube is buoyant if and only if
\(\frac{m_\text{env}}{m_\text{air}}<1\).
This is only fulfilled for a medium of rather hight density and low
pressure.
E.g. for \(E_\text{max}=\unit{670}{\mega\volt\per\meter}\),
\(\epsilon_r=2.3\),
\(\rho_\text{env}=\unit{0.94}{\gram\per\cubic{\centi\metre}}\)
\cite{WIKI-PE}
(for Polyethylen) and
\(\rho_\text{air}=\unit{1.204}{\kilogram\per\cubic\metre}\)
\cite{WIKI-AIR}
we get \(\frac{m_\text{env}}{m_\text{air}}> 35\).
\comment{ {bc -ql}-script
epsilon_r = 2.3
epsilon_0 = 8.854187817*10^-12
epsilon = epsilon_r * epsilon_0
p_air = 101325
rho_air = 1.204
rho_env = 0.94/1000*(100)^3
e_max = 670*10^6
"1/(4*p_air) * (epsilon * e_max^2 - 4*p_air + sqrt(epsilon^2*e_max^4-8*p_air*epsilon*e_max^2))"
x=1/(4*p_air) * (epsilon * e_max^2 - 4*p_air + sqrt(epsilon*e_max^2)*sqrt(epsilon*e_max^2-8*p_air))
.00004308749095311312
"(2*x+1)/(x^2+2*x+1) * rho_env/rho_air"
(2*x+1)/(x^2+2*x+1) * rho_env/rho_air
35.01567190569879919850
}

\section{Further arguments against free electron gas}
\label{sec:arguments}
Although we have proven already in section~\ref{sec:gauss} that
there is no free electron gas in the tube, we present here
additional arguments.

First, if an electron in the tube moves the charges on the envelope
reacts on this.
For a perfect conductor they react such that the total electrical
field outside the tube vanishes.
For a real conductor this is only partially achieved or rather with
a time delay.
Since the time of Sir Isaac Newton we know actio equals
reactio\cite[Lex. III., page 13]{Newton}.
Thus, the moving of the charges on the envelope causes forces on all
the electrons in the tube.
Therefore, the electrons are \emph{not free}.

Second, we consider the situation of conductors. Vacuum is a
conducter as soon as there are charge carriers in it, e.g. electrons.
By definition of vacuum there are no particles which hinder the
motion of the charge carriers but the carriers themselves.
Because of the electrostatic induction the mobile electrical charges
resides on the surface of the conductor. This is a well known fact
of electrostatic theory.
The electrons are not in the interior of the vacuum but at its
boundary, i.e. the surface \(S_1\).

Third, from particle accelerator experiments we know for sure that
we must contain an ensemble of electrons by some means, e.g.
magnetic fields or the self magnetic field if there is an electric
current in the vacuum. In the latter case the self magnetic field
reduces the radial pressure of the electrons.
Without containment the cloud of electrons expands until it reaches the
boundaries of the container.

\section{Conclusion}
The properties of the tube as described in
\cite{AB-Tube} are quite promising. Unfortunately our analysis shows that
such tubes are \emph{not} possible with today, i.e. 2011, known
materials. Thus we cannot build ballons with high lift capacity,
high altitude platforms, or quasi-superconductors as proposed in
\cite{AB-Tube} with this technique.

This analysis uses that the electrons inside the tube adhere as
static electricity to the envelope. The resulting pressure is that
of the electric field.

A.~Bolonkin's opinion is that most of the electrons are reflected by
the envelope and as such exchange momentum with it which creates
outward pressure. Other contributions of the pressure are
\enquote{charge pressure}, and \enquote{magnetic pressure}.
This effects are additional to the electric pressure.
\cite{AB-private}

We recommend an experiment to observe the real behaviour of the
electrons in the tube.

\appendix
\section{Material constants}
\label{ssec:material}
\begin{tabularx}{\textwidth}[t]{ll}
   Substance       &  Dielectric Strength [\(\mega\volt\per\meter\)]
   \\
   \hline
   Helium                &0.15 \\
   Air                   &0.4 - 3.0 \\
   Alumina               &13.4 \\
   Window glass          &9.8 - 13.8 \\
   Silicone oil          &10 - 15 \\
   Polystyrene           &19.7 \\
   Neoprene rubber       &15.7 - 27.6 \\
   Water                 &65 - 70 \\
   Salt                  &150 \\
   Benzene               &163 \\
   Teflon                &87 - 173 \\
   L.D. Polyethylene film&300 \\
   Fused silica          &470 - 670 \\
\end{tabularx}
\begin{flushright}
Source: \cite{WIKI-DE}
\end{flushright}

\end{document}